\documentclass[journal]{IEEEtran}
\usepackage{textcomp}
\usepackage{url}
\usepackage{verbatim}
\usepackage{etoolbox}
\usepackage{amsmath,amssymb}
\usepackage{graphicx,graphics,color,psfrag}
\usepackage{balance}

\allowdisplaybreaks[4]
\usepackage{algorithm}
\usepackage{accents}
\usepackage{amsthm}
\usepackage{bm}
\usepackage{algorithmic}
\usepackage[english]{babel}
\usepackage{multirow}
\usepackage{enumerate}
\usepackage{cases}
\usepackage{stfloats}
\usepackage{dsfont}
\usepackage{color,soul}
\usepackage{amsfonts}
\usepackage{graphicx,amsmath,amssymb}
\usepackage{fancyhdr}
\usepackage{hhline}
\usepackage{graphicx,graphics}
\usepackage{array,color}
\usepackage{amsmath}
\usepackage{float}
\usepackage{amssymb}
\usepackage{amsmath}
\usepackage{bbm}
\usepackage{amsthm}
\usepackage{amsfonts}
\usepackage{graphicx}
\usepackage{algorithm}
\usepackage{algorithmic}
\usepackage{epstopdf}
\usepackage{cite}
\usepackage{amsmath,bm}
\usepackage{subfigure}
\usepackage{graphicx}
\usepackage{color}
\usepackage{graphicx}
\usepackage{calc}
\usepackage{diagbox}

\begin{document}
\bibliographystyle{IEEEtran}

\title{Aggregation Design for Personalized Federated Multi-Modal Learning over Wireless Networks \thanks{All authors are with Cooperative Medianet Innovation Center and Shanghai Key Laboratory of Digital Media Processing and Transmission, Shanghai Jiao Tong University, Shanghai, China (e-mail: \{yinbsh, zhiyongchen, mxtao\}@sjtu.edu.cn). M. Tao is also with Department of Electronic Engineering, Shanghai Jiao Tong University, China. \emph{(Corresponding author: Zhiyong Chen, Meixia Tao)}}}

\author{Benshun Yin, Zhiyong Chen, \emph{Senior Member, IEEE}, and Meixia Tao, \emph{IEEE Fellow}\vspace{-2em}}
\maketitle

\begin{abstract}
Federated Multi-Modal Learning (FMML) is an emerging field that integrates information from different modalities in federated learning to improve the learning performance. In this letter, we develop a parameter scheduling scheme to improve personalized performance and communication efficiency in personalized FMML, considering the non-independent and non-identically distributed (non-IID) data along with the modality heterogeneity. Specifically, a learning-based approach is utilized to obtain the aggregation coefficients for parameters of different modalities on distinct devices. Based on the aggregation coefficients and channel state, a subset of parameters is scheduled to be uploaded to a server for each modality. Experimental results show that the proposed algorithm can effectively improve the personalized performance of FMML.
\end{abstract}
\begin{IEEEkeywords}
	Federated learning, Multi-Modal learning, Aggregation coefficients.
\end{IEEEkeywords}

\section{Introduction}
Federated learning (FL) \cite{federated,designfed} is a well-known distributed learning framework. In recent years, FL has evolved from focusing solely on single-modal data to incorporating multiple modalities, termed Federated Multi-Modal Learning (FMML) \cite{fed_mmwave,FedMSplit,contra_rep,fed_HAR}. By integrating information from different modalities, FMML facilitates more comprehensive representation of data, improving the accuracy and robustness of models. The key of FMML is leveraging multi-modal data on devices for collaborative learning, capitalizing on the synergistic potential of varied data types to improve learning outcomes.

\begin{figure}[t]
	\centering
	\includegraphics[width=7cm]{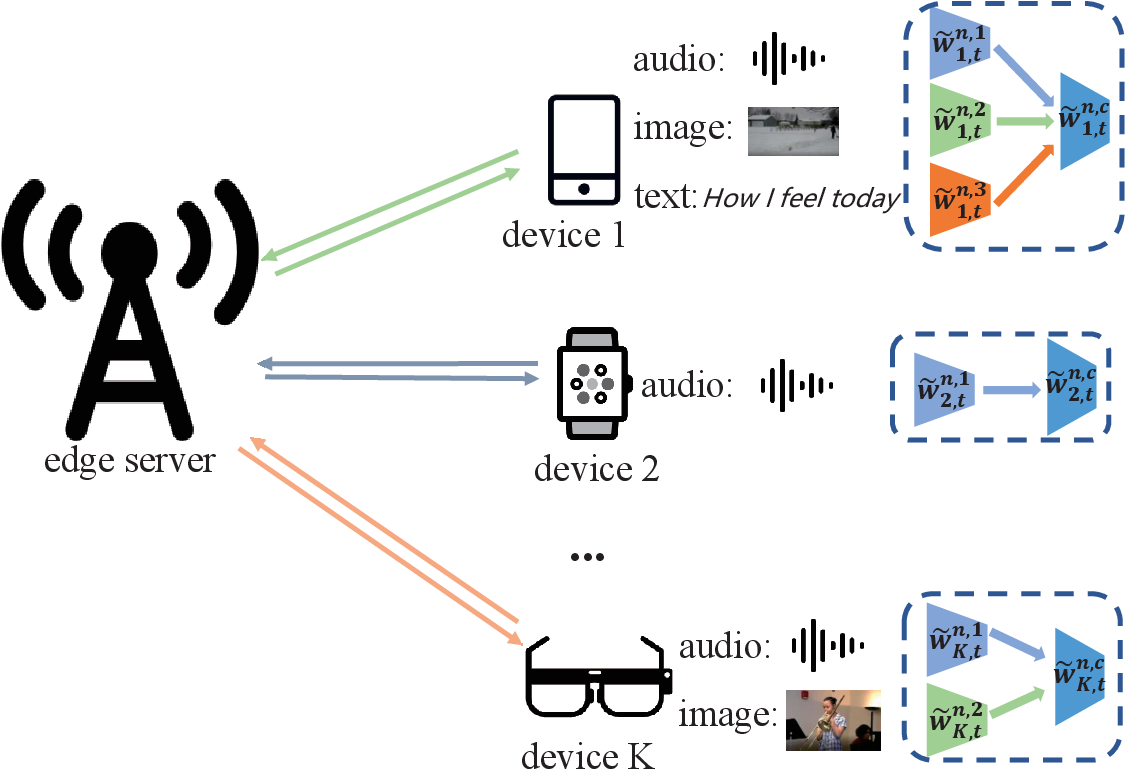}
	\caption{A federated multi-modal learning system.}
	\label{system}
\end{figure}

The data modalities on different devices can be \textbf{heterogeneous} in FMML, due to variations in detection environments and device types, as shown in Fig. \ref{system}. For example, some vehicles may be equipped with visual sensors only, whereas others possess both visual and radar sensors. Generally, distinct neural networks are employed to extract features from different modalities. For instance, the transformer \cite{transformer} can be used for processing text, whereas convolutional neural networks are applied for visual data. The modality heterogeneity in FMML implies that the parameters corresponding to the modalities a device possesses can be trained locally. On the other hand, the non-independent and identically distributed (\textbf{non-IID}) data across devices can lead to the local model converging towards personalized data distributions. To address the challenges posed by modality heterogeneity and non-IID data, we optimize the aggregation coefficients of each modality on each device in this letter. 


Many federated learning methods \cite{fedprox,fedfomo,device_sche,online_sche,fedAMP} have been proposed to address the challenge of non-IID data, but these methods are exclusively designed for single-modal data. Several recent works \cite{fed_mmwave,FedMSplit,contra_rep,fed_HAR,dmml-kd} have considered the multi-modal data in distributed learning. To address the issue of modality heterogeneity, the training scheme of FMML has been designed in \cite{FedMSplit,contra_rep,fed_HAR}. A dynamic and multi-view graph structure is applied on the edge server to automatically capture the relationships among devices in \cite{FedMSplit}. In \cite{contra_rep}, an inter-modal contrastive objective is designed to complement the absent modality. To learn the cross-modal features, the modality-agnostic features and the modality-specific features are extracted from each modality separately in \cite{fed_HAR}. However, existing works have not considered aggregating sub-networks from part of users for enhancing the personalized performance.


Inspired by the above, we improve the model aggregation process of FMML to enhance the personalized performance and communication efficiency. Firstly, we employ a learning-based approach to optimize the aggregation coefficients of different devices for each modality. The aggregation coefficients are updated using a gradient descent approach, which is seamlessly integrated into the FMML training process without introducing additional communication overhead. Secondly, we develop a parameter scheduling method to improve the communication efficiency based on the aggregation coefficients and channel conditions. Finally, we conduct experiments to demonstrate that the proposed approach can effectively improve the personalized performance of FMML.

\section{System Model}
\label{sys_model}

\subsection{Multi-Modal Data and Neural Networks}
We consider a federated multi-modal learning system as shown in Fig. \ref{system}, which consists of a set of wireless devices $\mathcal{K}=\{1,2,...,K\}$. The set of modalities of device $k$ is $\mathcal{M}_k\subseteq \mathcal{M}$, where $\mathcal{M}=\{1,2,...,M\}$ contains all modalities. Note that $\mathcal{M}_k$ varies across different devices. Each device $k\in \mathcal{K}$ has its local dataset $\{(\{\bm{x}^m_{k,1}\}_{m\in\mathcal{M}},y_{k,1}),...,(\{\bm{x}^m_{k,D_k}\}_{m\in\mathcal{M}},y_{k,D_k})\}$ with the size $D_k$. $\bm{x}^m_{k,1},...,\bm{x}^m_{k,D_k}$ are the raw data of the modality $m\in\mathcal{M}_k$. $y_{k,1},...,y_{k,D_k}$ refer to the corresponding labels. 

As shown in Fig. \ref{system}, the data $\bm{x}_{k,d}^m$ of modality $m$ is processed by the network with the parameter $\tilde{\bm{w}}^{n,m}_{k,t}$ to extract the feature. The features of different modalities are concatenated and then processed by the classifier with the parameter $\tilde{\bm{w}}^{n,c}_{k,t}$ to obtain the prediction. Using the output and the label, the loss function, such as cross-entropy, can be computed to evaluate the performance of multi-modal fusion. These parameters are obtained by device $k$ after the $n$-th iteration of the $t$-th global round. For convenience, we denote the parameter specific to modality $m$ as $\bm{w}^{n,m}_{k,t}$. Part of parameters in $\tilde{\bm{w}}^{n,c}_{k,t}$ are shared across different modalities, which is denoted as $\bm{w}^{n,M+1}_{k,t}$. In summary, the parameters that trained by device $k$ is $\bm{w}^{n}_{k,t}=\{\bm{w}^{n,m}_{k,t}\}_{m\in \mathcal{M}_k}\cup\{\bm{w}^{n,M+1}_{k,t}\}$.


\subsection{Personalized Federated Multi-Modal Learning}
In this paper, we aim to minimize the personalized loss function $F_k(\bm{w})=\frac{1}{D_k}\sum_{d=1}^{D_k}f(\{\bm{x}^m_{k,d}\}_{m\in\mathcal{M}_k},y_{k,d};\bm{w})$ for device $k$ through the collaborative learning of devices. Here, $f(\cdot)$ is the task-specific loss function, e.g., cross-entropy. The execution process of personalized FMML can be divided into two stages, i.e., local update and parameter aggregation. The local update and parameter aggregation stages are performed for many global rounds to obtain the desired learning performance. 


\subsubsection{Local Update Stage}
Device $k$ uses the parameter $\bm{w}^{n,m}_{k,t}$, which is specific to modality $m \in \mathcal{M}_k$, along with the shared parameter $\bm{w}^{n,M+1}_{k,t}$, to calculate the loss function. Then, the gradient is calculated to update the parameter $\bm{w}^{n}_{k,t}$. It's not necessary for device $k$ to maintain the parameters corresponding to the other modalities, i.e., $\bm{w}^{n,m}_{k,t}, m \notin \mathcal{M}_k$, because they are not used in the forward and backward propagation. The update equation of device $k$ is
\begin{align}
	\label{grad}
	\bm{w}^n_{k,t} =\bm{w}^{n-1}_{k,t}-\eta \nabla_{\bm{w}^{n-1}_{k,t}} F_k (\bm{w}^{n-1}_{k,t}),
\end{align}
for $n=1,2,...,N_k$. $N_k$ is the number of local iterations performed on device $k$. 

\subsubsection{Parameter Aggregation Stage}

We use $I_{k,t}^m=1$ denote that device $k$ uploads its locally trained parameter $\bm{w}^{N_k,m}_{k,t}$ to the server in the $t$-th round, otherwise $I_{k,t}^m=0$. In personalized FMML, instead of obtaining a global model through parameter aggregation, the server retains a personalized model $\bm{W}_{k,t}=\{\bm{W}^{m}_{k,t}\}_{m\in \mathcal{M}_k}\cup\{\bm{W}^{M+1}_{k,t}\}$ for each device. $\bm{W}^{m}_{k,t}$ and $\bm{W}^{M+1}_{k,t}$ represent the parameter specific to modality $m$ and the parameter shared across modalities respectively. For device $k$ and modality $m$ with $I_{k,t}^m=1$, the server-side parameter is replaced with the aggregated parameter, i.e.,
\begin{align}
	\label{avg}
	\bm{W}^m_{k,t}=\sum_{k^{'}=1}^K \xi^m_{k,k^{'},t} \bm{w}^{N_k,m}_{k^{'},t}, m=1,2,...,M+1,
\end{align}
where $\xi^m_{k,k^{'},t}\ge0, \forall k^{'}, \forall k$ are the aggregation coefficients for the parameters. For device $k$ and modality $m$ with $I_{k,t}^m=0$, $\xi^m_{k,k^{'},t}$ and $\xi^m_{k^{'},k,t}$ are set as 0 for $\forall k^{'}\neq k$. Besides, the coefficients satisfy $\sum_{k^{'}=1}^K\xi^m_{k,k^{'},t}=1$. The server-side parameter with $I_{k,t}^m=0$ remain unchanged, i.e., $\bm{W}^m_{k,t}=\bm{W}^m_{k,t-1}$. 

After parameter aggregation, the parameter $\bm{W}^m_{k,t}$ corresponding to $I_{k,t}^m=1$ is transmitted to device $k$. For device $k$ and modality $m$ with $I_{k,t}^m=1$, the initial parameter in the $(t+1)$-th round is $\bm{w}^{0,m}_{k,t+1}=\bm{W}^{m}_{k,t}$. For the parameters of the other sub-networks, the initial parameter in the $(t+1)$-th round is $\bm{w}^{0,m}_{k,t+1}=\bm{w}^{N_k,m}_{k,t}$. 

It can be observed from (\ref{avg}) that for modality $m$, only devices with $I_{k,t}^m=1$ upload the parameters to the server for aggregation. Additionally, due to distinct personalized objectives on the devices, the aggregation coefficients vary across different devices. Optimizing the aggregation coefficients for each device is necessary to facilitate parameter aggregation that benefits the enhancement of personalized performance. If the parameters from other devices are not very helpful to the performance improvement of modality $m$ on device $k$, device $k$ can choose not to upload the parameter specific to modality $m$ to the server for aggregation, thereby reducing communication overhead.

\section{Aggregation Design for Personalized Federated Multi-Modal Learning}
\label{prob}

\subsection{Learn to Update Aggregation Coefficients}
To update the aggregation coefficients for improving the personalized performance, we initially define an parameter matrix for each modality $m\in\mathcal{M}$, i.e., 
\begin{align}
	\bm{\Xi}^m_t=\begin{bmatrix}
		\hat{\xi}^m_{1,1,t}&...&\hat{\xi}^m_{1,K,t}\\
		...&\hat{\xi}^m_{k,k^{'},t}&...\\
		\hat{\xi}^m_{K,1,t}&...&\hat{\xi}^m_{K,K,t}\\
	\end{bmatrix}\in \mathbb{R}^{K\times K}.
\end{align}

Due to the unknown data similarity between devices at the initial stage, all values in $\bm{\Xi}^m_t$ are initialized to $\frac{1}{K}$. Then we transform these parameters to construct the aggregation coefficients that satisfy the requirements of being greater than 0 and summing to 1. Firstly, we apply the softmax function to the parameters $\hat{\xi}^m_{k,k^{'},t}$ for each modality of each device, i.e.,
\begin{align}
	\tilde{\xi}^m_{k,k^{'},t}=\frac{e^{\hat{\xi}^m_{k,k^{'},t}}}{\sum_{k^{''}=1}^{K}e^{\hat{\xi}^m_{k,k^{''},t}}}, \forall k^{'},\forall k,\forall m.
\end{align}
Considering that a subset of parameters is uploaded, we further employ $I_{k,t}^m$ to transform $\tilde{\xi}^m_{k,k^{'},t}$. We define a matrix $\tilde{\bm{I}}^m_{t}\in \mathbb{R}^{K\times K}$, where the $k$-th row and $k^{'}$-th column of $\tilde{\bm{I}}^m_{t}$ is $\tilde{I}^m_{k,k^{'},t}$. For device $k$ and modality $m$ with $I_{k,t}^m=0$, we set $\tilde{I}^m_{k,k,t}=1$ and $\tilde{I}^m_{k,k^{'},t}=\tilde{I}^m_{k^{'},k,t}=0$ for $k^{'}\neq k$. The other elements in $\tilde{\bm{I}}^m_{t}$ are set as 1. Then we use $\tilde{I}^m_{k,k^{'},t}$ to obtain $\xi^m_{k,k^{'},t}$, i.e.,
\begin{align}
	\xi^m_{k,k^{'},t}=\frac{\tilde{I}^m_{k,k^{'},t}\tilde{\xi}^m_{k,k^{'},t}}{\sum_{k^{''}=1}^{K}\tilde{I}^m_{k,k^{''},t}\tilde{\xi}^m_{k,k^{''},t}}, \forall k^{'},\forall k,\forall m.
\end{align}

The gradient of $\bm{\Xi}^m_t$ can be obtained using the chain rule. Suppose that device $k$ downloads $\bm{W}^m_{k,t}$ from the server as the initial parameter $\bm{w}^{0,m}_{k,t+1}$, then the gradient of $\hat{\xi}^m_{k,k^{'},t}$ is
\begin{align}
	\nabla_{\hat{\xi}^m_{k,k^{'},t}}F_k (\bm{W}^m_{k,t})= (\nabla_{\hat{\xi}^m_{k,k^{'},t}}\bm{W}^m_{k,t})^T\nabla_{\bm{W}^m_{k,t}} F_k (\bm{w}^{0}_{k,t+1}).
\end{align}
The server can obtain the gradient $\nabla_{\hat{\xi}^m_{k,k^{'},t}}\bm{W}^m_{k,t}$ after the parameter aggregation. The gradient $\nabla_{\bm{W}^m_{k,t}} F_k (\bm{w}^{0}_{k,t+1})$ can be estimated by the change of parameters. For example, if device $k$ uploads $\bm{w}^{N_k,m}_{k,t+1}$ to the server in the $(t+1)$-th round, the gradient $\nabla_{\bm{W}^m_{k,t}} F_k (\bm{w}^{0}_{k,t+1})$ can be estimated using $\bm{w}^{N_k,m}_{k,t+1}-\bm{W}^m_{k,t}$. The parameter $\hat{\xi}^m_{k,k^{'},t}$ is updated by 
\begin{align}
	\hat{\xi}^m_{k,k^{'},t+1}=\hat{\xi}^m_{k,k^{'},t}-\hat{\eta}\nabla_{\hat{\xi}^m_{k,k^{'},t}}F_k (\bm{W}^m_{k,t}),
\end{align}
where $\hat{\eta}$ is the learning rate.

\begin{algorithm}[t]
	\caption{Personalized FMML with the update of aggregation coefficients}
	\label{alg}
	\hspace*{0.02in}{\bf Input:}
	Initial parameters $\bm{W}_{k,0}$, $\bm{w}^0_{k,0}$ and $\Xi_t^m$; Learning rate $\eta$ and $\hat{\eta}$; Initial value $A^m_k=0$.
	
	\begin{algorithmic}[1]
		\FOR {training round $t=1$ to $\tau$}
		\FOR {all devices $k=1$ to $K$ in parallel}
		\STATE Local update for $N_k$ iterations with equation (\ref{grad}).
		\ENDFOR
		\FOR {modality $m=1$ to $M+1$}
		\STATE Sort $\frac{1-\tilde{\xi}^m_{k,k,t}}{T^{ParD}_{k,t}+T^{cmp}_{k,t}+\frac{\sum_{m^{'}=1}^{m-1} I^{m^{'}}_{k,t}\Upsilon^{m^{'}}+ \Upsilon^m}{\mathcal{B}\log(1+\frac{p_kg_{k,t}^2}{\mathcal{B}\mathcal{N}_0})}}$ for devices with modality $m$ in descending order.
		\STATE Select the top $\hat{K}$ devices with the largest values. Set $I^m_{k,t}=1$ and $A^m_k=0$ for them. Set $I^m_{k,t}=0$ and $A^m_k=A^m_k+1$ for the others.
		\STATE For device $k$ satisfying $A^m_k\ge A^{th}$ and $m\in\mathcal{M}_k$, set $I^m_{k,t}=1$ and $A^m_k=0$.
		\ENDFOR
		\STATE Transmit $\bm{w}^{N_k,m}_{k,t}$ with $I^m_{k,t}=1$ to the server.
		\STATE The server aggregates the received parameters with the coefficient $\xi^m_{k,k^{'},t}$, and retain the gradient $\nabla_{\hat{\xi}^m_{k,k^{'},t}}\bm{W}^m_{k,t}$.
		\STATE  The server calculate $\bm{w}^{N_k,m}_{k,t}-\bm{W}^m_{k,t-1}$ and combine it with $\nabla_{\hat{\xi}^m_{k,k^{'},t-1}}\bm{W}^m_{k,t-1}$ to obtain $\bm{\Xi}^m_{t+1}$.
		\STATE The server sends $\bm{W}^m_{k,t}$ with $I^m_{k,t}=1$ to the respective devices.
		\ENDFOR
	\end{algorithmic}
\end{algorithm}

\begin{figure*}[t]
	\centering
	\includegraphics[width=14cm]{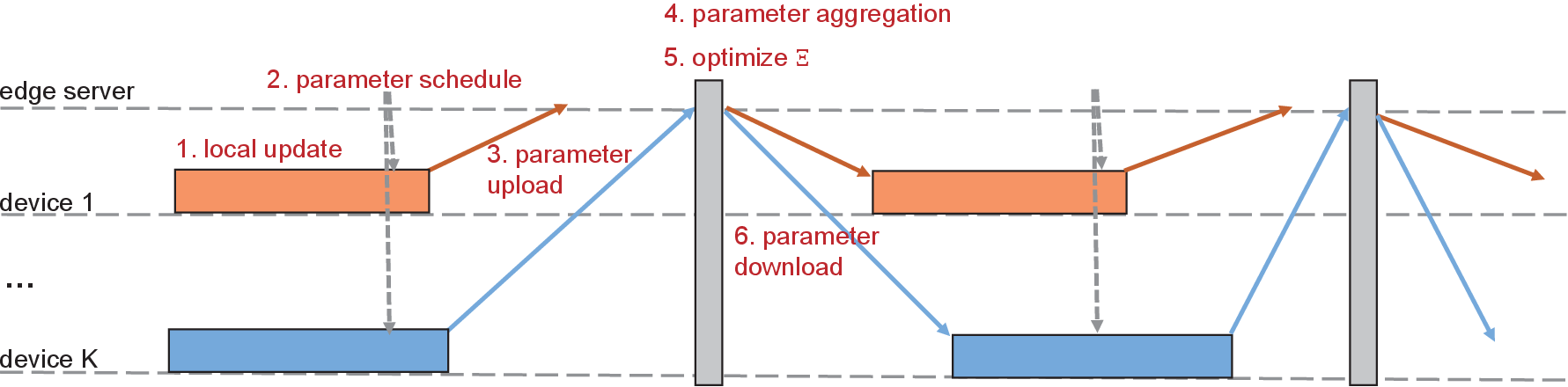}
	\caption{Execution process of personalized federated multi-modal learning systems with the update of aggregation coefficients.}
	\label{slot}
\end{figure*}
The execution process of personalized FMML with the update of aggregation coefficients is shown in Fig. \ref{slot}. Firstly, the devices perform local updates using (\ref{grad}) in the $t$-th global round. Secondly, the server schedules the parameters specific to modality $m$ from a subset of devices, which are then aggregated with others to improve performance. If the parameter specific to modality $m$ of device $k$ is scheduled, the index $I^m_{k,t}$ is set to 1. Otherwise $I^m_{k,t}=0$. After the server scheduling, device $k$ uploads the parameter $\bm{w}^{N_k,m}_{k,t}$ with $I^m_{k,t}=1$ to the server. Upon receiving the parameters, the server performs aggregation using the aggregation coefficients $\xi^m_{k,k^{'},t}$ to obtain $\bm{W}^m_{k,t}$. After the aggregation, the gradient $\nabla_{\hat{\xi}^m_{k,k^{'},t}}\bm{W}^m_{k,t}$ can be obtained and retained at the server for updating the coefficient in the next round. Then the parameter change $\bm{w}^{N_k,m}_{k,t}-\bm{W}^m_{k,t-1}$ is calculated with the uploaded parameter $\bm{w}^{N_k,m}_{k,t}$, which is combined with the gradient $\nabla_{\hat{\xi}^m_{k,k^{'},t-1}}\bm{W}^m_{k,t-1}$ obtained from the $(t-1)$-th round to update $\bm{\Xi}^m_t$, resulting in $\bm{\Xi}^m_{t+1}$. For device $k$ and modality $m$ with $I^m_{k,t}=0$, the parameter change is set as 0. Therefore, only the subset of elements in $\bm{\Xi}^m_t$ corresponding to $I^m_{k,t}=1$ is updated in the $t$-th round. Finally, the server-side parameter $\bm{W}^m_{k,t}$ specific to modality $m$ with $I^m_{k,t}=1$ is transmitted to device $k$. The overall training process of the proposed personalized FMML with the update of aggregation coefficients is outlined in Algorithm \ref{alg}.

\subsection{Improvement on Communication Efficiency}
To improve the communication efficiency, we schedule the parameters specific to modality $m$ from $\hat{K}$ devices. We use the obtained aggregation coefficients to determine the necessity of aggregating users' parameters within each modality. Specifically, a larger value of $\xi^m_{k,k^{'},t}$ indicates a greater necessity to aggregate the parameter $\bm{w}^{N_k,m}_{k,t}$ with $\bm{w}^{N_{k^{'}},m}_{k^{'},t}$ for improving the personalized performance on device $k$. $\sum_{k^{'}\neq k} \xi^m_{k^{'},k,t}=1-\xi^m_{k,k,t}$ represents the impact of $\bm{w}^{N_k,m}_{k,t}$ on the parameters specific to modality $m$ of the other devices. $\sum_{k^{'}\neq k} \xi^m_{k,k^{'},t}$ can represent the total impact of parameters $\bm{w}^{N_{k^{'}},m}_{k^{'},t}, k^{'}\neq k$ from other devices on the parameter $\bm{w}^{N_k,m}_{k,t}$ of device $k$. Because $\xi^m_{k,k^{'},t}$ and $\xi^m_{k^{'},k,t}$ are determined by the data similarity between device $k$ and $k^{'}$, the values of $\sum_{k^{'}\neq k} \xi^m_{k^{'},k,t}$ and $\sum_{k^{'}\neq k} \xi^m_{k,k^{'},t}$ are generally close. Therefore, devices with a larger value of $1-\xi^m_{k,k,t}$ are more likely to upload the parameter $\bm{w}^{N_k,m}_{k,t}$ to the server. 

Additionally, it's crucial to consider the impact of channel conditions. For computing the training latency conveniently, synchronous aggregation method is adopted as shown in Fig. \ref{slot}. Parameters are aggregated after receiving all the scheduled parameters. Therefore, when scheduling parameters, we aim for a shorter duration between two consecutive aggregation processes. In the $t$-th training round, the download latency of device $k$ is determined by the parameter scheduling $I^m_{k,t-1}$ in the $(t-1)$-th round, i.e., 
\begin{align}
	T^{ParD}_{k,t}= \frac{\sum_{m\in\mathcal{M}_k} I^m_{k,t-1}\Upsilon^m +I^{M+1}_{k,t-1}\Upsilon^{M+1}}{\mathcal{B}\log(1+\frac{\hat{p}g_{k,t}^2}{\mathcal{B}\mathcal{N}_0})},
\end{align}
where $\Upsilon^m$ denotes the parameter size (\emph{in bit}) specific to modality $m$ and $\Upsilon^{M+1}$ is the size of the shared parameters. The server transmits the parameters to device $k$ using a channel with a bandwidth of $\mathcal{B}$ through the transmission power $\hat{p}$. The channel gain between device $k$ and server in the $t$-th round is $g_{k,t}$. The spectral density of the additive white Gaussian noise (AWGN) is $\mathcal{N}_0$. The latency of local update is
\begin{align}
	T^{cmp}_{k,t}= \frac{N_k(\sum_{m\in\mathcal{M}_k} O^m +O^{M+1})}{f_ke_k},
\end{align}
where $O^m$ is the floating point operations (FLOPs) specific to modality $m$ in each iteration and $O^{M+1}$ is the FLOPs for the shared parameters. $f_k$ and $e_k$ are the computing frequency and the number of FLOPs per cycle for device $k$, respectively. 

When scheduling parameters, it is necessary to consider the completion time of local updates on devices. If $T^{ParD}_{k,t}+T^{cmp}_{k,t}$ is large, uploading numerous parameters of device $k$ may extend the duration of the training round. To balance personalized performance and latency, we adopt the parameter scheduling method outlined in lines 5 to 9 of Algorithm \ref{alg} to select $\hat{K}$ devices. Specifically, $p_k$ is the transmission power of device $k$. Device $k$ has not uploaded the parameters of modality $m$ to the server for $A_k^m$ rounds. If $A_k^m$ exceeds the threshold $A^{th}$, device $k$ must upload the parameters once to prevent insufficient optimization of the corresponding aggregation coefficients.


\section{Simulation Results}
\label{sim}

\subsection{Simulation Setup}
In the simulation, we consdier a FMML system, where devices are uniformly distributed within a circular area with a diameter of 100 meters. The distance between device $k$ and base station is $d_{k}$. The channel gain $g_{k,t}$ follows the Rayleigh distribution with the mean $10^{-PL(d_{k})/20}$. The path loss is $PL(d_{k})(dB)=32.4+20\log_{10}(\hat{f}_k^{carrier})+20\log_{10}(d_{k,k^{'}})$ and the carrier frequency is $\hat{f}_k^{carrier}=2.6$ GHz. 

Two datasets are utlitzed in the simulation. The CREMA-D dataset \cite{cremad} comprises data from six categories, featuring both visual and audio modalities. This dataset is distributed across 9 devices, with one-third of these devices processing both modalities and the remaining devices processing only one modality. Two ResNet-18 \cite{resnet} with different input dimensions are used to extract features of length 512 from the $257\times 188$ audio inputs and $224\times 224 \times3$ visual inputs, respectively. The extracted features are then concatenated and fed into a fully connected neural network with node sizes of [1024, 1024, 6]. The MOSEI dataset \cite{mosei} comprises data from seven categories, encompassing visual, audio, and textual modalities. It is partitioned among 18 devices, where one-third of the devices have all three modalities, another third have two modalities, and the final third have only one modality. The input feature dimensions for audio, visual and text are 74, 35, 300, respectively. Three different transformers are applied to extract features of length 128 by processing the inputs, respectively. The extracted features are then concatenated and fed into a fully connected neural network with node sizes of [384, 384, 7]. The training is performed for 50 rounds with $\eta=2\times 10^{-4}$, $\hat{\eta}=0.01$ and $A^{th}=10$.

We apply three different distributions of data categories. \textbf{Non-IID-1}: each device only possesses data from any three categories. \textbf{Non-IID-2}: 50\% data on each device belongs to one category, while the remaining data is randomly selected. \textbf{Non-IID-3}: 30\% data on each device belongs to one category, while the remaining data is randomly selected.

\begin{table}[t]
	\centering
	\caption{Accuracies on CREMA-D with $\hat{K}=\frac{K}{3}$}
	\begin{tabular}{cccc}
		\hline
		& non-IID-1 & non-IID-2 & non-IID-3\\
		\hline
		FedAvg&48.90 &52.99  &36.90 \\
		Local training&54.93 &53.51 &39.70 \\
		FedProx&51.02 &53.08 &36.90 \\		
		FedFomo&57.53 &54.03 &43.14 \\		
		FedAMP&58.65 &53.90 &41.91 \\		
		Proposed&\textbf{65.31} &\textbf{55.98} &\textbf{48.11} \\
		\hline
	\end{tabular}
	\label{acc_cremad}
\end{table}

\begin{table}[t]
	\centering
	\caption{Accuracies on MOSEI with $\hat{K}=\frac{K}{3}$}
	\begin{tabular}{cccc}
		\hline
		& non-IID-1 & non-IID-2 & non-IID-3\\
		\hline
		FedAvg&49.13 &52.46  &43.54 \\
		Local training&50.81 &52.72 &43.21 \\
		FedProx&49.05 &53.01 &43.27 \\		
		FedFomo&50.88 &53.44 &43.81 \\
		FedAMP&50.15 &53.28 &43.65 \\				
		Proposed&\textbf{52.88} &\textbf{54.81} &\textbf{45.47} \\
		\hline
	\end{tabular}
	\label{acc_mosei}
\end{table}

\subsection{Performance Comparison}
The test accuracies of the proposed method are compared with those of FedAvg \cite{federated}, local training, FedProx \cite{fedprox}, FedFomo \cite{fedfomo} and FedAMP \cite{fedAMP}. As shown in Tables \ref{acc_cremad} and \ref{acc_mosei}, the proposed method achieves the highest test accuracies. For example, in the Non-IID 1 case of CREMA-D, the accuracy can be increased from 50.63\% to 60.49\%. Compared to FedAvg and FedProx, which train a global model, local training focuses solely on the performance of each device, exhibiting better personalized performance. FedFomo is specifically designed to enhance the personalized performance of devices. The proposed method outperforms FedFomo, indicating that it can effectively update the aggregation coefficients.

In addition to the ratio shown in line 6 of Algorithm \ref{alg}, we also utilize the linear combination $1-\tilde{\xi}^m_{k,k,t}-\alpha\big(T^{ParD}_{k,t}+T^{cmp}_{k,t}+\frac{\sum_{m^{'}=1}^{m-1} I^{m^{'}}_{k,t}\Upsilon^{m^{'}}+ \Upsilon^m}{\mathcal{B}\log(1+\frac{p_kg_{k,t}^2}{\mathcal{B}\mathcal{N}_0})}\big)$ as a metric for scheduling, where $\alpha$ is a hyperparameter. The accuracy and training latency of the two methods are shown in Table \ref{sche_method}. Increasing $\alpha$ can bias the scheduling method towards reducing latency, while causing the decrease in accuracy. Compared to the linear combination, scheduling using the ratio can achieve higher accuracy within the same training time. Table \ref{up_devices} shows the performance under different $\hat{K}$. We can observe an overall improvement in performance with the increase of $\hat{K}$. 


Table \ref{time_save} presents the training time of different methods. The proposed method, by considering the impact of parameter scheduling on each round's duration, achieves lower latency than FedAvg. In contrast, FedFomo, which necessitates downloading all parameters uploaded by other devices for local testing, incurs the longest training time.

\begin{table}[t]
	\centering
	\caption{Comparison of different scheduling methods}
	\begin{tabular}{ccccc}
		\hline
		Method &ratio  &$\alpha=10^{-4}$ & $\alpha=10^{-3}$ & $\alpha=10^{-2}$ \\
		\hline
		Accuracy& 65.31 & 65.32 & 64.52 & 63.21\\
		Training time&5.30$\times 10^3$ & 5.47$\times 10^3$& 5.39$\times 10^3$ & 4.68$\times 10^3$\\
		\hline
	\end{tabular}
	\label{sche_method}
\end{table}

\begin{table}[t]
	\centering
	\caption{Accuracies on CREMA-D with different $\hat{K}$}
	\begin{tabular}{ccccc}
		\hline
		& & $\hat{K}=\frac{K}{3}$ & $\hat{K}=\frac{2K}{3}$ & $\hat{K}=K$\\
		\hline
		\multirow{3}{*}{non-IID-1}&FedAvg&48.90 &49.02  &48.83 \\
		\multirow{3}{*}{}		&FedAMP&58.65 &58.55 &59.48 \\		
		\multirow{3}{*}{}		&Proposed&65.31 &64.18 &\textbf{66.89} \\
		\hline
		\multirow{3}{*}{non-IID-2}&FedAvg&52.99 &54.86  &54.10 \\
		\multirow{3}{*}{}		&FedAMP&53.90 &55.53 &55.07 \\		
		\multirow{3}{*}{}		&Proposed&55.98 &58.56 &\textbf{59.74} \\
		\hline
		\multirow{3}{*}{non-IID-3}&FedAvg&36.90 &40.38  &41.05 \\
		\multirow{3}{*}{}		&FedAMP&41.91 &40.26 &41.12 \\		
		\multirow{3}{*}{}		&Proposed&48.11 &47.98 &\textbf{48.13} \\
		\hline
	\end{tabular}
	\label{up_devices}
\end{table}

\begin{table}[t]
	\centering
	\caption{Training time (Seconds) of different algorithms}
	\begin{tabular}{cccc}
		\hline
		&$\hat{K}=\frac{K}{3}$  &$\hat{K}=\frac{2K}{3}$  &$\hat{K}=K$ \\
		\hline
		FedAvg&6.50$\times 10^3$ &7.47$\times 10^3$ & 7.87$\times 10^3$  \\
		FedFomo&7.95$\times 10^3$ & 1.13$\times 10^4$& 1.46$\times 10^4$ \\
		Proposed& 5.47$\times 10^3$&7.10$\times 10^3$& 7.87$\times 10^3$ \\
		\hline
	\end{tabular}
	\label{time_save}
\end{table}

\begin{figure}[t]
	\centering
	\subfigure[]{
		\includegraphics[width=4cm]{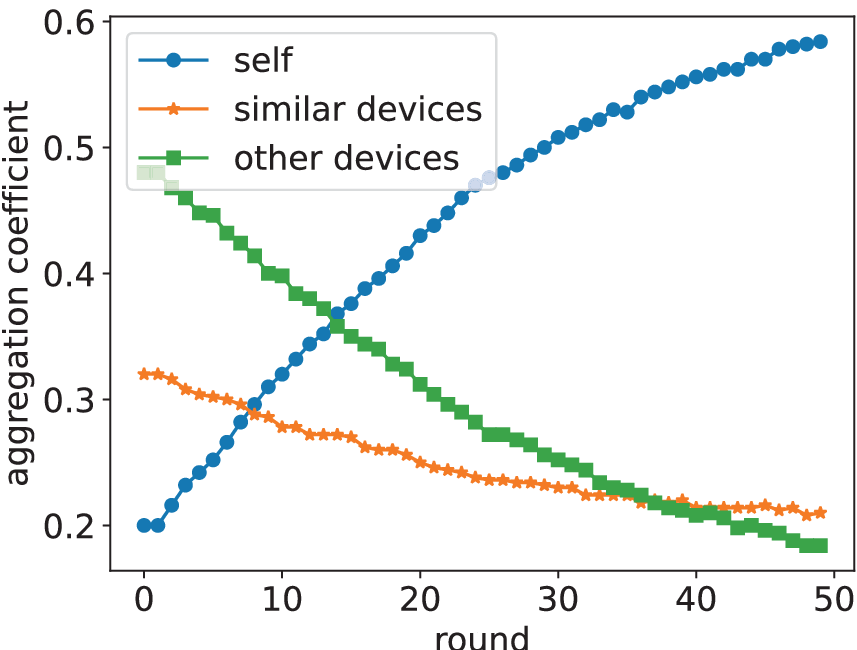}
	}
	\subfigure[]{
		\includegraphics[width=4cm]{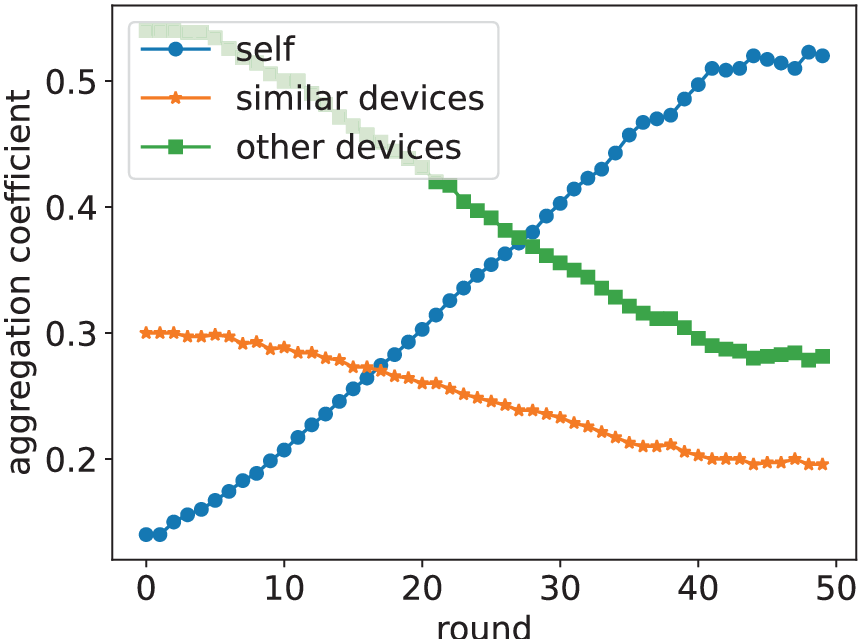}
	}
	\caption{The variation of the aggregation coefficients of (a) audio modality, (b) visual modality on CREMA-D with the non-IID-1 distribution.}
	\label{weight_cur}
\end{figure}

Fig. \ref{weight_cur} illustrates the variation of aggregation coefficients for both modalities on CREMAD with the non-IID-1 distribution. It is observed that the aggregation coefficients for each device's own parameters increase significantly. Meanwhile, the aggregation coefficients for devices with a similar data distribution, characterized by having two common categories, gradually decrease. In cases where the similarity in data distribution among devices is very low (either lacking identical categories or having only one common category), the aggregation coefficients for these devices decrease significantly.

\section{Conclusion}
\label{con}
To improve personalized performance in FMML, this letter adopts a learning-based approach to obtain the aggregation coefficients for parameters across various modalities on distinct devices. For improving the communication efficiency, we further design a parameter scheduling method, taking into account both the aggregation coefficients and the channel state of devices. Experimental results show that the proposed method effectively improve the personalized performance of FMML, notably reducing training time.

%

\bibliography{reference}
	
\end{document}